\documentclass[12pt]{article}
\usepackage{epsf}
\title{ {\bf
The flavor changing $t\rightarrow c \, l_1^- l_2^+$ decay in the general two 
Higgs doublet model.}}
\author{\vspace{1cm}\\
        {\bf E. O. Iltan}
        \thanks{E-mail address:
        eiltan@heraklit.physics.metu.edu.tr}\,\,\, and 
   {\bf I. Turan}
        \thanks{E-mail address:
        ituran@.metu.edu.tr}
 \\
        Physics Department, Middle East Technical University \\
        Ankara, Turkey\\}
\date{}

\begin{document}
\setlength{\baselineskip}{24pt}
\maketitle
\setlength{\baselineskip}{7mm}
\begin{abstract}
We study the flavor changing $t\rightarrow c \, l_1^- l_2^+$ decay in the 
framework of the general two Higgs doublet model, so called model III. 
We predict the branching ratio for $l_1=\tau,\, l_2=\mu$ at the order of 
the magnitude of $BR \sim 10^{-8}$.
\end{abstract} 
\thispagestyle{empty}
\newpage
\setcounter{page}{1}
\section{Introduction}
The top quark has a large mass and therefore it breaks the $SU(2)\times U(1)$ 
symmetry maximally.  Richness of the decay products stimulates one to
study its decays to test the standard model (SM) and to get some clues about 
the new physics, beyond. The rare decays of the top quark have been studied 
in the literature, in the framework of the SM and beyond \cite{Mahlon}-
\cite{IltanH0}; the one-loop flavor changing transitions $t\rightarrow c g 
(\gamma, Z)$ in \cite{Liu,Eilam3}, $t\rightarrow c V 
(V V)$ in \cite{Liu2} and $t\rightarrow c H^0$ in \cite{Eilam, Eilam3, 
Mele, Tao,IltanH0}. 

These decays are strongly suppressed in the SM and the predicted values  
of the branching ratio ($BR$) of the process $t\rightarrow c g(\gamma, 
Z)$ is  $4 \times 10^{-11}\, (5 \times 10^{-13},\, 1.3 \times 10^{-13}\,)$ 
\cite{Eilam}, the $BR$ for $t\rightarrow c H^0$ is  at the order 
of the magnitude of $10^{-14}-10^{-13}$, in the SM \cite{Mele}. These
prediction are so small that it is not possible to measure them even 
at the highest luminosity accelerators. This forces one to go beyond the SM
and study these rare decays in the framework of new physics.   
$t\rightarrow c H^0$ decay has been studied in the general two Higgs doublet 
model (model III) \cite{IltanH0} and it has been found that the  $BR$ of this 
process could reach to the values of order $10^{-6}$, playing with the free 
parameters of the model III, respecting the existing experimental 
restrictions. This is a strong enhancement, almost seven orders larger 
compared to the one in the SM. 

The present work is devoted to the analysis of the flavor changing (FC)
$t\rightarrow c \, (l_1^- l_2^+ + l_1^+ l_2^-)$ decay in the framework 
of the general two Higgs doublet model (model III). This decay occurs 
in the tree level since the FC transitions in the quark and leptonic sector 
are permitted in the model III. Here, the Yukawa couplings for $t-c$ and 
$l_1^- - l_2^+$ transitions play the main role and they exist with the help 
of the internal neutral Higgs bosons, $h^0$  and $A^0$. In the process, it is 
possible to get $h^0$ and $A^0$ resonances since the kinematical region is 
large enough and this difficulty can be solved by choosing the appropriate 
propagator for $h^0$ and $A^0$ (see section 2). In the tree level, the $BR$ of 
the $t\rightarrow c \, (l_1^- l_2^+ + l_1^+ l_2^-)$ decay  for $l_1=\tau$ and 
$l_2=\mu$ is predicted as $10^{-8}-10^{-7}$. We also calculate the one loop 
effects related with the interactions due to the internal mediating charged 
Higgs boson (see Fig. \ref{fig1}: b,c,d) and observe that their contribution 
to the $BR$ is negligible, namely $10^{-11}-10^{-10}$. 

The paper is organized as follows:
In Section 2, we present the $BR$ of the decay $t\rightarrow c
\, (l_1^- l_2^+ + l_1^+ l_2^-)$ in the framework of model III. Section 3 is 
devoted to discussion and our conclusions.

\section{The flavor changing $t\rightarrow c\, (l_1^- l_2^+ + l_1^+ l_2^-)$ 
decay in the framework of the general two Higgs Doublet model} 
The flavor changing transition $t\rightarrow c \, l_1^- l_2^+ $ is forbidden 
in the SM. Such transitions would be possible in the case that the Higgs 
sector is extended and the flavor changing neutral currents (FCNC) in the 
tree level are permitted. The simplest model which obeys these features is 
the model III version of the two Higgs doublet model (2HDM). This section 
is devoted to the calculation of the $BR$, in the model III. In this model, 
there are various new parameters, such as complex Yukawa couplings, masses 
of new Higgs bosons., etc... and  they should be restricted by using the 
present experimental results.

The $t\rightarrow c \, l_1^- l_2^+ $ process is controlled by the Yukawa 
interaction and, in the model III, it reads 
\begin{eqnarray}
{\cal{L}}_{Y}&=&\eta^{U}_{ij} \bar{Q}_{i L} \tilde{\phi_{1}} U_{j R}+
\eta^{D}_{ij} \bar{Q}_{i L} \phi_{1} D_{j R}+
\xi^{U\,\dagger}_{ij} \bar{Q}_{i L} \tilde{\phi_{2}} U_{j R}+
\xi^{D}_{ij} \bar{Q}_{i L} \phi_{2} D_{j R} 
\nonumber \\ &+& 
\eta^{E}_{ij} \bar{l}_{i L} \phi_{1} E_{j R}+
\xi^{E}_{ij} \bar{l}_{i L} \phi_{2} E_{j R} + h.c. \,\,\, ,
\label{lagrangian}
\end{eqnarray}
where $L$ and $R$ denote chiral projections $L(R)=1/2(1\mp \gamma_5)$, 
$\phi_{i}$ for $i=1,2$, are the two scalar doublets, $\bar{Q}_{i L}$ are 
left handed quark doublets, $U_{j R} (D_{j R})$ are  right handed up (down) 
quark singlets, $l_{i L}$ ($E_{j R}$) are lepton doublets (singlets), with 
family indices $i,j$. The Yukawa matrices $\xi^{U,D}_{ij}$ and 
$\xi^{E}_{ij}$ have in general complex entries. It is possible to collect 
SM particles in the first doublet and new particles in the second one by 
choosing the parametrization for $\phi_{1}$ and $\phi_{2}$ as
\begin{eqnarray}
\phi_{1}=\frac{1}{\sqrt{2}}\left[\left(\begin{array}{c c} 
0\\v+H^{0}\end{array}\right)\; + \left(\begin{array}{c c} 
\sqrt{2} \chi^{+}\\ i \chi^{0}\end{array}\right) \right]\, ; \, 
\phi_{2}=\frac{1}{\sqrt{2}}\left(\begin{array}{c c} 
\sqrt{2} H^{+}\\ H_1+i H_2 \end{array}\right) \,\, .
\label{choice}
\end{eqnarray}
with the vacuum expectation values,  
\begin{eqnarray}
<\phi_{1}>=\frac{1}{\sqrt{2}}\left(\begin{array}{c c} 
0\\v\end{array}\right) \,  \, ; 
<\phi_{2}>=0 \,\, ,
\label{choice2}
\end{eqnarray}
and considering the gauge and $CP$ invariant Higgs potential which 
spontaneously breaks  $SU(2)\times U(1)$ down to $U(1)$  as:
\begin{eqnarray}
V(\phi_1, \phi_2,\phi_3 )&=&c_1 (\phi_1^+ \phi_1-v^2/2)^2+
c_2 (\phi_2^+ \phi_2)^2 \nonumber \\ &+& +
c_3 [(\phi_1^+ \phi_1-v^2/2)+ \phi_2^+ \phi_2]^2
+ c_4 [(\phi_1^+ \phi_1) (\phi_2^+ \phi_2)-(\phi_1^+ \phi_2)(\phi_2^+ \phi_1)]
\nonumber \\ &+& 
c_5 [Re(\phi_1^+ \phi_2)]^2 +
c_{6} [Im(\phi_1^+ \phi_2)]^2 
+c_{7} \, ,
\label{potential}
\end{eqnarray}
with constants $c_i, \, i=1,...,7$. Here, $H_1$ and $H_2$ are the mass 
eigenstates $h^0$ and $A^0$ respectively, since no mixing occurs between 
two CP-even neutral bosons $H^0$ and $h^0$ in the tree level, for our choice. 

The Flavor Changing (FC) interaction can be obtained as 
\begin{eqnarray}
{\cal{L}}_{Y,FC}=
\xi^{U\,\dagger}_{ij} \bar{Q}_{i L} \tilde{\phi_{2}} U_{j R}+
\xi^{D}_{ij} \bar{Q}_{i L} \phi_{2} D_{j R} + 
\xi^{E}_{ij} \bar{l}_{i L} \phi_{2} E_{j R} + h.c. \,\, ,
\label{lagrangianFC}
\end{eqnarray}
where the couplings  $\xi^{U,D}$ for the FC charged interactions are
\begin{eqnarray}
\xi^{U}_{ch}&=& \xi^U_{N} \,\, V_{CKM} \nonumber \,\, ,\\
\xi^{D}_{ch}&=& V_{CKM} \,\, \xi^{D}_{N} \,\, ,
\label{ksi1} 
\end{eqnarray}
and $\xi^{U,D}_{N}$ is defined by the expression 
\begin{eqnarray}
\xi^{U (D)}_{N}=(V_{R (L)}^{U (D)})^{-1} \xi^{U,(D)} V_{L(R)}^{U (D)}\,\, .
\label{ksineut}
\end{eqnarray}
Here the index "N" in $\xi^{U,D}_{N}$ denotes the word "neutral". 
Notice that, in the following, we replace $\xi^{U,D,E}$ with 
$\xi^{U,D,E}_{N}$ where "N" denotes the word "neutral" and define 
$\bar{\xi}^{U,D,E}_N$ which satisfies the equation $\xi^{U,D,E}_N=
\sqrt{\frac{4\,G_F} {\sqrt{2}}}\, \bar{\xi}^{U,D,E}_N$.

\begin{figure}[htb]
\vskip -0.45truein
\centering
\epsfxsize=5.8in
\leavevmode\epsffile{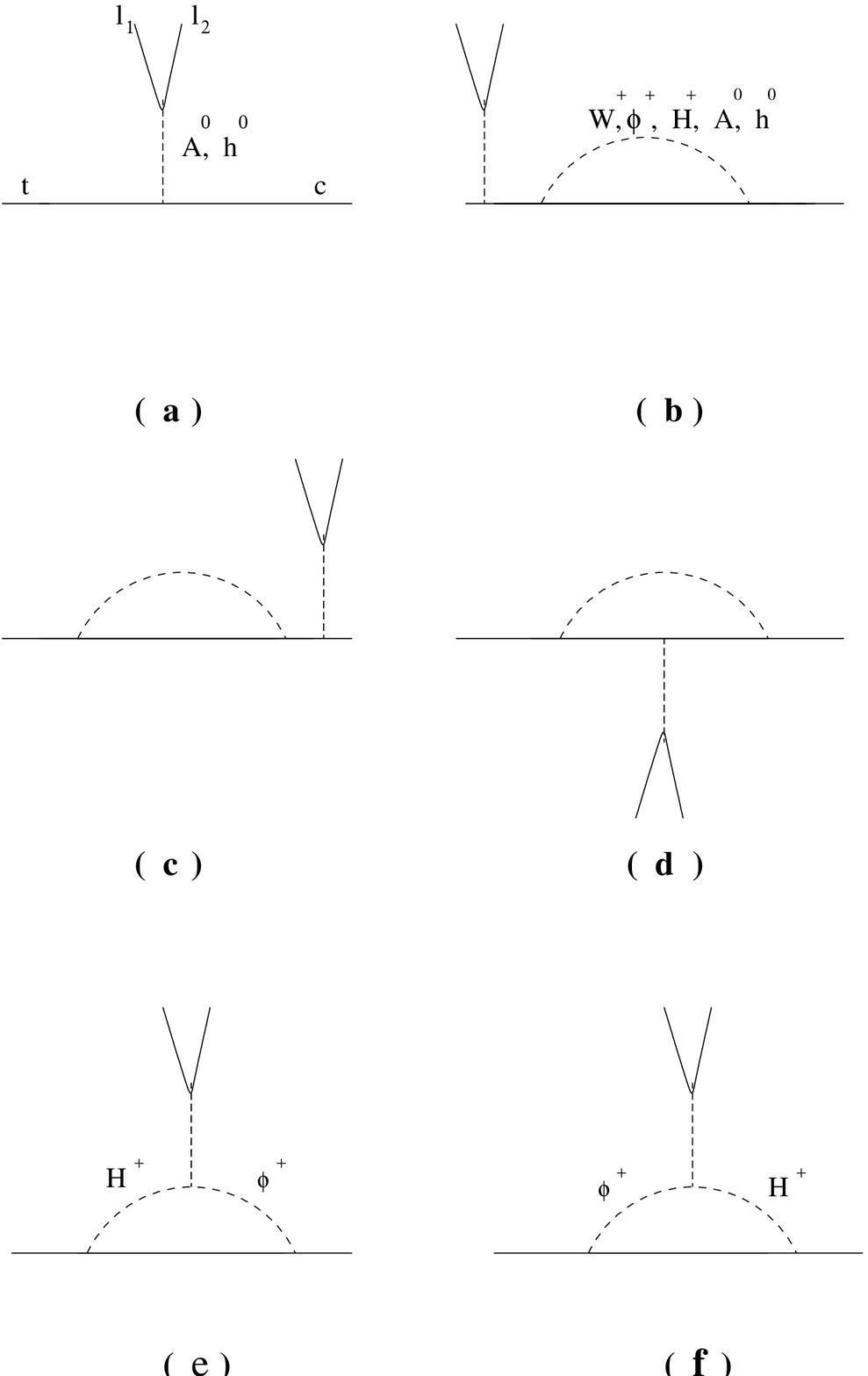}
\vskip -0.05truein
\caption[]{Tree level and one loop level diagrams contribute to the decay 
$t\rightarrow c \, l_1^-  l_2^+$. Dashed lines represent the 
$h^0,A^0, \phi^{\pm},W^{\pm}, \,H^{\pm}$ fields.}
\label{fig1}
\end{figure}

In the model III, the $t\rightarrow c \, l_1^- l_2^+ $ decay exists in the 
tree level, by taking non-zero $t-c\, (l_1^- - l_2^+)$ transition with the 
help of the neutral bosons $h^0$ and $A^0$. For completeness, we also take 
the one loop contributions into account (see Fig. \ref{fig1}) and, we use 
the onshell renormalization scheme to get rid of the existing divergences. 
The method is to obtain the renormalized $t\rightarrow c h^{0\, *} 
(A^{0\, *})$ transition vertex function 
\begin{eqnarray}
\Gamma_{REN}^{h^{0 *}}&=&\Gamma_{0}^{h^{0 *}}+\Gamma_{C}^{h^0} 
\, , \nonumber \\
\Gamma_{REN}^{A^{0 *}}&=&\Gamma_{0}^{A^{0 *}}+\Gamma_{C}^{A^0}
\, ,\label{GammaRen}
\end{eqnarray}
by using 
\begin{eqnarray}
\Gamma_{REN}^{h^0}|_{onshell}&=& \frac{i}{2 \sqrt{2}} \, \Big(
\,(\xi^{U}_{N, tc}+\xi^{U *}_{N, ct})+
(\xi^{U}_{N, tc}-\xi^{U *}_{N, ct}) \gamma_5\, \Big) 
\nonumber \\
\Gamma_{REN}^{A^0}|_{onshell}&=& -\frac{1}{2 \sqrt{2}} \, \Big(
\, (\xi^{U}_{N, tc}-\xi^{U *}_{N, ct})+
(\xi^{U}_{N, tc}+\xi^{U *}_{N, ct}) \gamma_5\, \Big)
\label{renorm}
\end{eqnarray}
and the counter term
\begin{eqnarray}
\Gamma_{C}^{h^0}&=& \Gamma_{REN}^{h^0}|_{onshell}-\Gamma_{0}^{h^0}|_{onshell}
\, , \nonumber \\
\Gamma_{C}^{A^0}&=& \Gamma_{REN}^{A^0}|_{onshell}-\Gamma_{0}^{A^0}|_{onshell} 
\, .
\label{counter}
\end{eqnarray}
where $\Gamma_{0}^{h^0}$ is the bare vertex function. Here, we take the
loop diagrams (see Fig. \ref{fig1}) including $H^{\pm}$ intermediate 
boson for FC interaction (Fig. \ref{fig1}: b,c,d) in the quark sector, since 
$\xi^{D}_{N, bb}$ and $\xi^{U}_{N, tt}$ are dominant couplings 
in the loop effects. Therefore, we neglect all the Yukawa couplings except 
$\xi^{D}_{N,bb}$ and $\xi^{U}_{N,tt}$ in the loop contributions.
Notice that the self energy diagrams do not give any contribution in the 
onshell renormalization scheme.  

The renormalized vertex function is connected to the $l_1^- l_2^+$ out going
leptons by intermediate $h^0$ and $A^0$ bosons as shown in the Fig.
\ref{fig1} and for the matrix element square of the process 
$t\rightarrow c\, (l_1^- l_2^+ + l_1^+ l_2^-)$ we get 
\begin{eqnarray}
|M|^2&=& 8\,m_t^2\, (1-s) \, \sum_{S=h^0,A^0} |p_S|^2\, 
\Big( |a^{(q)}_{S}|^2+|a^{\prime \, (q)}_{S}|^2 \Big)\, 
\Big( (s\, m_t^2 - (m_{l_1^-}-m_{l_2^+})^2 )\, |a^{(l)}_{S}|^2
\nonumber \\ &+&
(s\, m_t^2 - (m_{l_1^-}+m_{l_2^+})^2)\, |a^{\prime \, (l)}_{S}|^2 \Big) \, 
\nonumber \\
&+& 16\, m_t^2\, (1-s) \,\Bigg( 
(s\, m_t^2 - (m_{l_1^-}-m_{l_2^+})^2)\,  
Re [p_{h^0}\,p^*_{A^0}\, a^{(l)}_{h^0}\, a^{*(l)}_{A^0}\, 
(a^{(q)}_{h^0}\, a^{* (q)}_{A^0}+a^{\prime \, (q)}_{h^0}\, 
a^{\prime \,  * (q)}_{A^0})]
\nonumber \\ &+&
(s\, m_t^2 - (m_{l_1^-}+m_{l_2^+})^2)\, 
Re [p_{h^0}\,p^*_{A^0}\, a^{\prime \, (l)}_{h^0}\, a^{\prime \, * (l)}_{A^0}\, 
(a^{(q)}_{h^0}\, a^{* (q)}_{A^0}+
a^{\prime \, (q)}_{h^0}\, a^{\prime \, * (q)}_{A^0})] \Bigg ) \, ,
\label{M2tot}
\end{eqnarray}
where  
\begin{eqnarray}
p_S=\frac{i}{s\, m_t^2-m^2_S+i m_S\,\Gamma^{S}_{tot}} \, ,
\label{pS}
\end{eqnarray}
$\Gamma^{S}_{tot}$ is the total decay width of $S$ boson, for $S=h^0\,A^0$. 
Here, the parameter $s$ is $s=\frac{q^2}{m_t^2}$, 
and $q^2$ is the intermediate S boson momentum square. In eq. (\ref{M2tot}) 
the functions $a^{(l)}_{h^0, A^0}$, $a^{\prime \,(l)}_{h^0, A^0}$ have tree 
level contributions and $a^{(q)}_{h^0, A^0}$, $a^{\prime \, (q)}_{h^0, A^0}$ 
are the combinations of tree level and one-loop level contributions, 
\begin{eqnarray}
a^{(l)}_{h^0, A^0}&=&a^{Tree\, (l)}_{h^0, A^0}
\, , \nonumber \\  
a^{(q)}_{h^0, A^0}&=&a^{Tree\, (q)}_{h^0, A^0}+
a^{Loop\,(q)}_{h^0, A^0} 
\, , \nonumber \\ 
a^{\prime \, (l)}_{h^0, A^0}&=&a^{\prime\,  Tree\, (l)}_{h^0, A^0}
\, , \nonumber  \\ 
a^{\prime \, (q)}_{h^0, A^0}&=&a^{\prime\,  Tree\, (q)}_{h^0, A^0}+
a^{\prime \, Loop\,(q)}_{h^0, A^0} \,
\label{aql2}
\end{eqnarray}
and they read  
\begin{eqnarray}
a^{Tree\, (l)}_{h^0}&=&-\frac{i}{2\sqrt{2}}\, 
(\xi^E_{N,l_1 l_2}+\xi^{* E}_{N,l_2 l_1})
\, , \nonumber  \\ 
a^{Tree\, (l)}_{A^0}&=&\frac{1}{2\sqrt{2}}\, 
(\xi^E_{N,l_1 l_2}-\xi^{* E}_{N,l_2 l_1})
\, , \nonumber  \\ 
a^{\prime\, Tree\, (l)}_{h^0}&=&-\frac{i}{2\sqrt{2}}\, 
(\xi^E_{N,l_1 l_2}-\xi^{* E}_{N,l_2 l_1})
\, , \nonumber  \\ 
a^{\prime\, Tree\, (l)}_{A^0}&=&\frac{1}{2\sqrt{2}}\, 
(\xi^E_{N,l_1 l_2}+\xi^{* E}_{N,l_2 l_1})
\, , \nonumber  \\ 
a^{Tree\, (q)}_{h^0}&=&\frac{i}{2\sqrt{2}}\, 
(\xi^U_{N,tc}+\xi^{* U}_{N,ct})
\, , \nonumber  \\ 
a^{Tree\, (q)}_{A^0}&=&-\frac{1}{2\sqrt{2}}\, 
(\xi^U_{N,tc}-\xi^{* U}_{N,ct})
\, , \nonumber  \\ 
a^{\prime\, Tree\, (q)}_{h^0}&=&\frac{i}{2\sqrt{2}}\, 
(\xi^U_{N,tc}-\xi^{* U}_{N,ct})
\, , \nonumber  \\ 
a^{\prime\, Tree\, (q)}_{A^0}&=&-\frac{1}{2\sqrt{2}}\, 
(\xi^U_{N,tc}+\xi^{* U}_{N,ct})
\, , \nonumber  \\ 
a^{Loop\, (q)}_{h^0}&=&-\frac{i}{32\, \sqrt{2}\, \pi^2}\, V_{cb}\,V_{tb}^*
\xi^D_{N,bb}\, \Bigg( m_b^2\, \xi^D_{N,bb}\, 
\xi^{U *}_{N,tt} \, \int_0^1 dx \int_{0}^{1-x} dy \, f^{h^0}_1 (x,y) 
\nonumber \\ &+& 
m_b\, m_t\, (\xi^{D *}_{N,bb})^2\, 
\int_0^1 dx \int_{0}^{1-x} dy \, \Big( (1-x-y) \,f^{h^0}_1 (x,y) \Big) 
\nonumber \\ &-& 
m_b\, m_t\, |\xi^{D}_{N,bb}|^2\, 
\int_0^1 dx \int_{0}^{1-x} dy \, \Big( (x+y) \,f_1^{h^0} (x,y) \Big)  
\nonumber \\ &-&
\xi^{D *}_{N,bb}\, \xi^{U *}_{N,tt}\,
\int_0^1 dx \int_{0}^{1-x} dy \,f_2^{h^0} (x,y) \Bigg) 
\, , \nonumber \\ 
a^{Loop\, (q)}_{A^0}&=&\frac{1}{32\, \sqrt{2}\, \pi^2}\, V_{cb}\,V_{tb}^*
\, \xi^D_{N,bb}\, \Bigg( m_b^2\, \xi^D_{N,bb}\, 
\xi^{U *}_{N,tt} \, \int_0^1 dx \int_{0}^{1-x} dy f_1^{A^0} (x,y) 
\nonumber \\ &-& 
m_b\, m_t\, (\xi^{D *}_{N,bb})^2 \, 
\int_0^1 dx \int_{0}^{1-x} dy \, \Big( (1-x-y) \,f_1^{A^0} (x,y) \Big) 
\nonumber \\ &-&
m_b\, m_t\, |\xi^{D}_{N,bb}|^2\, 
\int_0^1 dx \int_{0}^{1-x} dy \,\Big( (x+y) \,f_1^{A^0} (x,y) \Big)  
\nonumber \\ &+&
\xi^{D *}_{N,bb}\, \xi^{U *}_{N,tt}\,
\int_0^1 dx \int_{0}^{1-x} dy \,f_2^{A^0} (x,y) \Bigg) 
\, , \nonumber \\ 
a^{\prime \, Loop\, (q)}_{h^0}&=&a^{Loop\, (q)}_{h^0}
\, , \nonumber \\ 
a^{\prime \, Loop\, (q)}_{A^0}&=&a^{Loop\, (q)}_{A^0}\, ,  
\label{aql3}
\end{eqnarray}
where 
\begin{eqnarray}
f_1^S&=&\frac{1}{L^S (m_S)}-\frac{1}{L^S (s)} \, , \nonumber \\ 
f_2^S&=& (1-x-y)\, (\frac{m_t^2\,x+m_S^2\,y)}{L^S (m_S)}-
\frac{m_t^2\,(x+s\,y)}{L^S (s)}) + 2\,ln \frac{L^S (s)}{L^S (m_S)} \, ,
\label{f12}
\end{eqnarray}
with 
\begin{eqnarray}
L^S (s)&=&m_b^2\,(x-1)+m^2_{H^{\pm}} x+m_t^2\,(-1+x+y)\,(x+s\,y)
\, , \nonumber \\ 
L^S (m_S)&=&m_b^2\,(x-1)+m^2_{H^{\pm}} x+(-1+x+y)\,(m_t^2\, x+m_S^2\,y)
\, . 
\label{LS}
\end{eqnarray}

Finally, the differential decay width (dDW)  
$\frac{d\Gamma}{ds}(t\rightarrow c \, (l_1^- l_2^++l_1^+ l_2^-) )$ is 
obtained by using the expression 
\begin{eqnarray}
\frac{d\Gamma}{ds}=\frac{1}{256\,N_c\,\pi^3}\,\lambda\, |M|^2 \, ,
\end{eqnarray}
where $\lambda$ is: \\
$\lambda=\frac{\sqrt{\Big(m_t^2\,(s-1)^2-4\,m_c^2 \Big) 
\, \Big( m_c^4+m_{l_1}^4+(m_{l_2}^2-m_t^2\,s)^2-2\,m_c^2\,(m_{l_1}^2+
m_{l_2}^2-m_t^2\,s)-2\,m_{l_1}^2\,(m_{l_2}^2+m_t^2\,s)  \Big)}}
{2\,m_t^2\,s}$. Here the parameter $s$ is restricted into the region 
$\frac{(m_{l_1}+m_{l_2})^2}{m_t^2}\leq s \leq \frac{(m_t-m_c)^2}{m_t^2}$. 
Notice that we use the parametrization $\xi^{E}_{N,l_1 l_2}=
|\xi^{E}_{N,l_1 l_2}|\, e^{i\theta_{l_1 l_2}}$ for the leptonic part, in the
numerical calculations.   
\section{Discussion}
This section is devoted to the analyses of the differential $BR$ ($dBR$) and  
the $BR$ of the process $t\rightarrow c\, (l_1^- l_2^+ + l_1^+ l_2^-)$ in the 
tree level and also in the one loop level, in the model III. The Yukawa 
couplings  $\xi^{U}_{N,tc}$ and $\xi^{E}_{N,l_1 l_2}$ play the main role in the tree level and new
couplings, especially $\xi^{D}_{N,bb}$, $\xi^{U}_{N,tt}$, enter 
into calculations if one goes to the loop level. Since these couplings are
free parameters of the model used, it is necessary to restrict them, using
appropriate experimental results. We use the constraint region 
by restricting the Wilson coefficient $C_7^{eff}$, which is the effective 
coefficient of the operator $O_7 = \frac{e}{16 \pi^2} \bar{s}_{\alpha} 
\sigma_{\mu \nu} (m_b R + m_s L) b_{\alpha} {\cal{F}}^{\mu \nu}$
(see \cite{Alil1} and references therein), in the region 
$0.257 \leq |C_7^{eff}| \leq 0.439$. Here upper and lower limits were 
calculated using the CLEO measurement \cite{cleo2}
\begin{eqnarray}
BR (B\rightarrow X_s\gamma)= (3.15\pm 0.35\pm 0.32)\, 10^{-4} \,\, ,
\label{br2}
\end{eqnarray}
and all possible uncertainities in the calculation of $C_7^{eff}$ 
\cite{Alil1}. The above restriction ensures to get upper and lower limits
for $\xi^{D}_{N,bb}$, $\xi^{U}_{N,tt}$ and also for $\xi^{U}_{N,tc}$ 
(see \cite{Alil1} for details). In our numerical calculations we choose the 
upper limit for $C_7^{eff}>0$, fix $\xi^{D}_{N,bb}=30\,m_b$ and take  
$\xi^{U}_{N,tc}\sim 0.01\, \xi^{U}_{N,tt}\sim 0.0025$, respecting the
constraints mentioned. Furthermore, the couplings $\xi^{E}_{N,l_1 l_2}$ 
in the leptonic part are restricted by using the experimental results, such
as, anomalous magnetic moment of muon, dipole moments of leptons, rare 
leptonic decays. For $l_1=\tau$ and $l_2=\mu$, we take the upper limit 
obtained by using experimental result of anomalous magnetic moment of muon 
\cite{Erilano}. For $l_1=\tau$ and $l_2=e$, we use the numerical result 
obtained for the couplings $\xi^{E}_{N,\tau e}$ in \cite{Erillep}, based on 
the experimental measurement of the leptonic process $\mu\rightarrow e\gamma$ 
\cite{MEGA}. The total decay widths of $h^0$ and $A^0$ are unknown parameters 
and we expect that they are at the same order of magnitude of 
$\Gamma^{H^0}_{tot} \sim (0.1-1.0)\, GeV$, where $H^0$ is the SM Higgs boson. 
Notice that, we take the value of the total decay width 
$\Gamma_T \sim \Gamma (t\rightarrow b W)$ as $\Gamma_T=1.55\, GeV $ and
choose the numerical values $m_{h^0}=80\, GeV$ and $m_{A^0}=90\, GeV$, for 
the calculation of the $BR$.

\begin{figure}[htb]
\vskip -3.0truein
\centering
\epsfxsize=6.8in
\leavevmode\epsffile{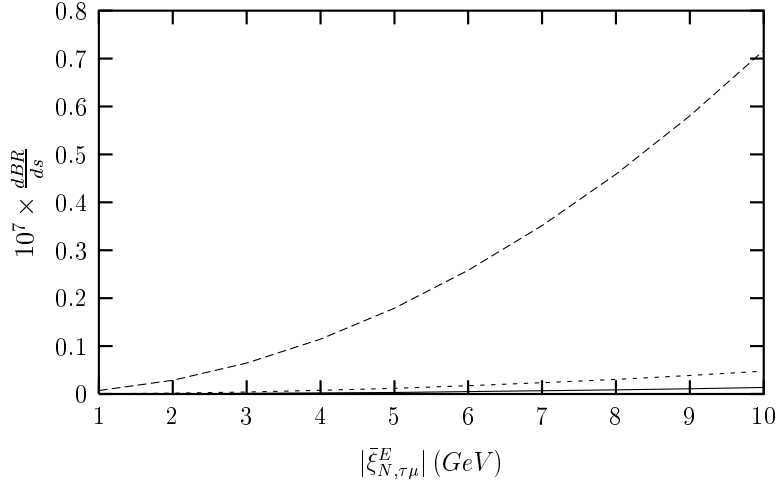}
\vskip -3.0truein
\caption[]{dBR $(t\rightarrow c\, (\tau^- \mu^+ + \tau^+ \mu^-))$ as a 
function of $|\bar{\xi}_{N, \tau\mu}^{E}|$ for $m_{h^0}=80\, GeV$, 
$m_{A^0}=90\, GeV$, $sin\,\theta_{\tau\mu}=0.5$, real $\bar{\xi}_{N,tc}^{U}$ 
and $\Gamma^{h^0}_{tot}=\Gamma^{A^0}_{tot}= 0.1\, GeV$. The solid (dashed, 
dash-dotted) line represents the case for $s=(\frac{10}{175})^2 
((\frac{50}{175})^2, (\frac{150}{175})^2)$.} 
\label{dBRsc1mutautree1050150}
\end{figure}

\begin{figure}[htb]
\vskip -3.0truein
\centering
\epsfxsize=6.8in
\leavevmode\epsffile{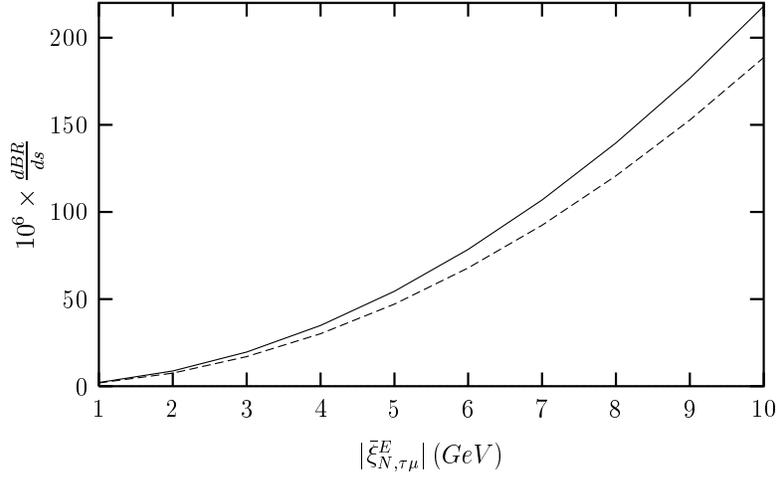}
\vskip -3.0truein
\caption[]{The same as Fig. \ref{dBRsc1mutautree1050150} but for 
$s=(\frac{80}{175})^2$ and $((\frac{90}{175})^2$. The solid (dashed)
line represents the case for $s=(\frac{80}{175})^2 ((\frac{90}{175})^2)$.}
\label{dBRsc1mutautree8090}
\end{figure}

In Fig. \ref{dBRsc1mutautree1050150}, we plot the dBR for the 
$t\rightarrow c\, (\tau^- \mu^+ + \tau^+ \mu^-)$ decay with respect to 
$|\bar{\xi}_{N, \tau\mu}^{E}|$ for $sin\,\theta_{\tau\mu}=0.5$, different 
$s$ values, $s=(\frac{10}{175})^2, (\frac{50}{175})^2, 
s=(\frac{150}{175})^2$. Here, we choose $\bar{\xi}_{N,tc}^{U}$ real and 
$\Gamma^{h^0}_{tot}=\Gamma^{A^0}_{tot}= 0.1\, GeV$. 
The solid (dashed, small dashed) line represents the case for 
$s=(\frac{10}{175})^2 ((\frac{50}{175})^2, (\frac{150}{175})^2)$.
From the figure, it is seen that the dBR is at the order of the magnitude of
$10^{-8}$ for $s=(\frac{50}{175})^2$ and 
$|\bar{\xi}_{N, \tau\mu}^{E}|\sim  5\, GeV$. dBR is less than $10^{-8}$ 
for $s=(\frac{10}{175})^2$ and $s=(\frac{150}{175})^2$ and it reaches 
extremely small values for $|\bar{\xi}_{N, \tau\mu}^{E}|\leq 1\, GeV$. 
Increasing  $|\bar{\xi}_{N, \tau\mu}^{E}|$ causes to enhance the dBR, as
expected. Fig. \ref{dBRsc1mutautree8090} is devoted to the same
dependence for $s=(\frac{80}{175})^2$ (solid line), $(\frac{90}{175})^2$ 
(dashed line), where the values of $s$ are taken at the $h^0$ and $A^0$ 
resonances.  The dBR is at the order of the magnitude of $10^{-6}$ for 
the small values of the coupling $|\bar{\xi}_{N, \tau\mu}^{E}|$ and
increases extremely with the increasing values of this coupling.   

\begin{figure}[htb]
\vskip -3.0truein
\centering
\epsfxsize=6.8in
\leavevmode\epsffile{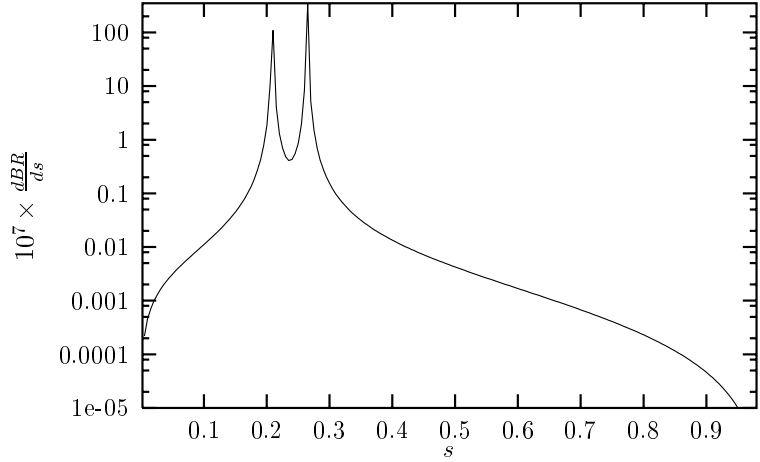}
\vskip -3.0truein
\caption[]{dBR $(t\rightarrow c\, (\tau^- \mu^+ + \tau^+ \mu^-))$ as a 
function of $s$ for $m_{h^0}=80\, GeV$, $m_{A^0}=90\, GeV$,
$|\bar{\xi}_{N, \tau\mu}^{E}|=10\, GeV$, 
$sin\,\theta_{\tau\mu}=0.5$, real $\bar{\xi}_{N,tc}^{U}$ and 
$\Gamma^{h^0}_{tot}=\Gamma^{A^0}_{tot}= 0.1\, GeV$.} 
\label{dBRsmutautrees}
\end{figure}

In Fig. \ref{dBRsmutautrees}, we plot the dBR with respect to $s$, for 
$|\bar{\xi}_{N, \tau\mu}^{E}|=10\, GeV$, $sin\,\theta_{\tau\mu}=0.5$ and 
$\Gamma^{h^0}_{tot}=\Gamma^{A^0}_{tot}= 0.1\, GeV$. It is observed that 
dBR has a strong $s$ dependence.

\begin{figure}[htb]
\vskip -3.0truein
\centering
\epsfxsize=6.8in
\leavevmode\epsffile{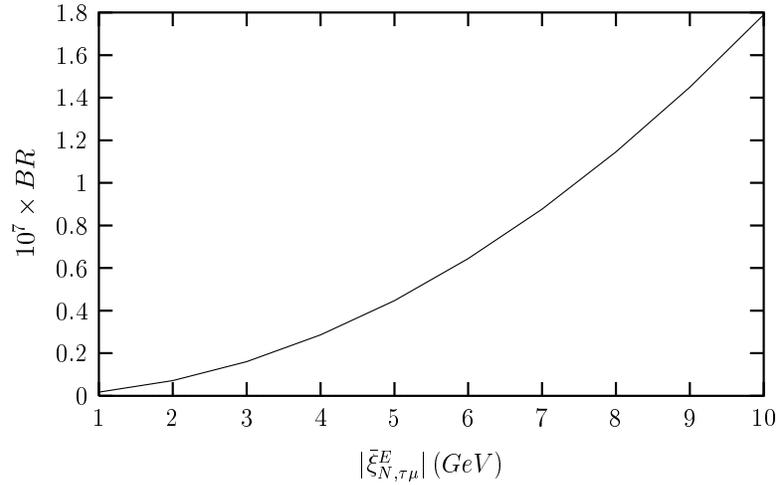}
\vskip -3.0truein
\caption[]{$BR\, (t\rightarrow c\, (\tau^- \mu^+ + \tau^+ \mu^-))$ as a 
function of $|\bar{\xi}_{N, \tau\mu}^{E}|$ for $m_{h^0}=80\, GeV$, 
$m_{A^0}=90\, GeV$, $sin\,\theta_{\tau\mu}=0.5$, 
real $\bar{\xi}_{N,tc}^{U}$ and $\Gamma^{h^0}_{tot}=\Gamma^{A^0}_{tot}= 
0.1\, GeV$.} 
\label{BRmutautree}
\end{figure}

Finally, in Fig. \ref{BRmutautree} we present the $BR$ for the process 
$t\rightarrow c\, (\tau^- \mu^+ + \tau^+ \mu^-)$ with respect to 
$|\bar{\xi}_{N, \tau\mu}^{E}|$ for $sin\,\theta_{\tau\mu}=0.5$ and 
$\Gamma^{h^0}_{tot}=\Gamma^{A^0}_{tot}= 0.1\, GeV$. The $BR$ is at the order 
of the magnitude of $10^{-8}$ for $|\bar{\xi}_{N, \tau\mu}^{E}|\sim 2 \, 
(GeV)$ and increases to the values $10^{-7}$ with increasing 
$|\bar{\xi}_{N, \tau\mu}^{E}|$. Notice that the one loop effects are at 
the order of the magnitude of $0.1\,\%$ of the tree level result and 
therefore their contribution is negligible.    

In the case of outgoing $\tau$ and $e$ leptons, the $BR$ is predicted at the
order of the magnitude of $10^{-14}-10^{-15}$, respecting the numerical
values of the coupling $|\bar{\xi}_{N, \tau e}^{E}|=(10^{-4}-10^{-3})\, 
GeV$, obtained in \cite{Erillep}, based on the experimental measurement of the 
leptonic process $\mu\rightarrow e\gamma$. For the outgoing $\mu$ and $e$ 
leptons, we believe that the $BR$ is extremely small, too difficult to be
measured.

At this stage we would like to summarize our results:

\begin{itemize}

\item The $BR$ of the flavor changing process 
$t\rightarrow c \, (l_1^- l_2^+ + l_1^+ l_2^-)$ is forbidden in the SM and 
the extended Higgs sector can bring considerable contribution to the $BR$ 
in the tree level, at the order of the magnitude of $10^{-8}-10^{-7}$, for 
$l_1=\tau$ and $l_1=\mu$. A measurement of such a $BR$ will be highly 
non-trivial due to efficiency problems in measuring the $\tau$-lepton and 
in identifying a c-quark jet. Moreover, one will have to overcome the problem 
of isolating the signal from possibly large reducible background by applying 
clever kinematical cuts which will further degrade the signal. 
However, the possible enhancement of the $BR$ of the given 
process in the model III forces one to search new models to get a measurable  
$BR$ theoretically. The $BR$ is sensitive to Yukawa coupling 
$\xi^E_{N,l_1 l_2}$ and, respecting the experimental limits on the relevant 
couplings, this results in extremely smaller $BR$'s of  
$t\rightarrow c \, (l_1^- l_2^+ + l_1^+ l_2^-)$, for $l_1=\tau,\, l_2=e$ and 
$l_1=\mu,\, l_2=e$, compared to the one for $l_1=\tau,\, l_2=\mu$. 
Notice that the loop effects are negligibly small.
\end{itemize}

Therefore, the future theoretical and experimental investigations of the 
process $t\rightarrow c  \, (l_1^- l_2^+ + l_1^+ l_2^-)$, especially for 
$l_1=\tau,\, l_2=\mu$, would play an important role in the determination 
the physics beyond the SM.
\section{Acknowledgement}
This work has been supported by the Turkish Academy of Sciences, in
the framework of the Young Scientist Award Program. 
(EOI/TUBA-GEBIP/2001-1-8). The authors would like to thank Professor 
T. M. Aliev for useful discussions.
\newpage
\newpage
\end{document}